\documentclass[aps,prb,preprint]{revtex4}

\begin{document}


\title{A Bell-type theorem without hidden variables}
\author{Henry P. Stapp}
\affiliation{Lawrence Berkeley National Laboratory,
University of California,
Berkeley, California 94720}

\begin{abstract}
It is shown that no theory that satisfies certain premises can
exclude faster-than-light influences. The premises include neither
the existence of hidden variables nor counterfactual definiteness,
nor any premise  that effectively entails the general existence of
outcomes of unperformed  local measurements. All the premises are
compatible with Copenhagen  philosophy and the principles and
predictions of relativistic quantum  field theory. The present
proof is contrasted with an earlier one with  the same objective.  
\end{abstract}

\maketitle

\section{Introduction}

The premises of Bell's original hidden-variable theorem\cite{1}
postulate  the existence of a substructure that determines in a
local manner the  outcomes of a set of alternative possible
measurements at most one of  which can actually be performed. The
implementation of the locality  condition in this way thus involves
a technical hidden-variable  assumption that goes beyond the
locality condition itself. Consequently,  Bell's proof of the
inconsistency of this local hidden-variable assumption  with
certain predictions of quantum theory casts no serious doubt on
the  locality condition: its implementation via the assumed
hidden-variable  substructure would appear to be the more likely
cause of the derived  inconsistency.  

Bell\cite{2} later introduced  a seemingly weaker local
hidden-variable assumption. However, this latter form can be
shown\cite{3,4} to be equivalent to the original one, apart from
errors that tend to zero as the number of  experiments tends to
infinity. Thus both forms of the hidden-variable assumption place
strong conditions on the class of theories  that are covered by the
theorems. These conditions are   logically equivalent to the
assumption that values can be pre-assigned  conjunctively and
locally to all of the outcomes of all of the alternative  possible
measurements. That assumption conflicts with what I
believe to  be the orthodox quantum philosophical attitude that one
should not make any assumption that effectively postulates the
existence of a well defined outcome of a localized measurement
process that is not performed. Thus these hidden-variable theorems
place in no serious jeopardy  the basic locality idea that a free
choice made by an experimenter  in one space-time region has no
influence in a second region that is  space-like separated from the
first. 

The present paper shows that this locality idea fails, however, not 
only under the hidden-variable assumption (or some closely related
presumption that effectively ensures the existence, within the
theory, of outcomes of  unperformed local measurements), but also in
a much larger class of theories,  namely those that are compatible
with the properties of free choice and  no backward-in-time
influence on observed outcomes, and that yield certain  predictions
of quantum theory in experiments of the Hardy type.\cite{5} The
first  two of these three properties is now described.

\smallskip \noindent  Free Choices.
{\it For the purposes of understanding and applying  quantum
theory, the choice of which experiment is to be performed in a 
certain space-time region can be treated as an independent free
variable  localized in that region.} Bohr repeatedly stressed the
freedom of the  experimenter to choose between alternative possible
options. This  availability of options is closely connected to his
complementarity  idea that the quantum state contains
complementary kinds of information  pertaining to the various
alternative mutually exclusive experiments  that might be chosen.
Of course, no two mutually incompatible  measurements can both be
performed, and an outcome of an experiment can  be specified only
under the condition that that particular experiment  be performed.
This ``free choice'' assumption is important because it allows the
causal part of cause-and-effect relationships to be identified:
{\it it  allows the choices made by experimenters to be considered
to be causes}. This identification underlies all Bell-type
arguments about causal relationships.

\smallskip \noindent No Backward-in-Time Influence (NBITI).
{\it An outcome that has already been observed and recorded in some
spacetime region at an earlier time can be considered fixed and settled, 
independently of which experiment a far-away experimenter will freely 
choose to perform at some later time.} This assumption assigns 
no value to a local measurement except under the condition that this
local measurement be performed. But any such locally observed value
is  asserted to be independent of which measurement will at some
later time be freely chosen and performed in a spacelike separated
region. 
This NBITI assumption is required to hold in at least one Lorentz
frame of reference, henceforth called LF. 

This NBITI assumption is compatible with relativistic quantum field theory.
In the Tomonaga-Schwinger\cite{6, 7} formulation the evolving state
is defined on a forward moving space-like surface.  Their work
shows that this surface can be defined in a continuum of ways
without altering the  predictions of the theory, so that no Lorentz
frame is  singled out as preferred. On the other hand, their
formalism  allows the quantum state to be defined by  the
constant time surfaces  in any one single Lorentz frame that one
wishes to choose, and shows that in this one frame the evolution,
including all reductions associated with specific outcomes of
measurements, proceeds forward in time, with a well defined past
that is not influenced either by later free choices  made by
experimenters or by the outcomes of the later measurements. Thus
this NBITI assumption is compatible with the principles and
predictions of relativistic quantum field theory. (Included in this
assumption is the tacit assumption that if an outcome appears to an
observer, then the mutually exclusive alternative does not occur:
the many-worlds idea that both outcomes occur is excluded.) 

This NBITI assumption is a small part of the larger locality
condition in  question here, which is the demand that what an
experimenter freely  chooses to do in one region has no effect in a
second region that is spacelike separated from the first. The no
backward-in-time part of the no-faster-than-light condition can be
imposed without generating any difficulties or conflict with
relativistic quantum field theory. But  a faster-than-light effect
then appears elsewhere. In particular the following theorem holds:\\

\noindent {\it Theorem}. Suppose a theory or model is compatible
with  the premises:

\begin{enumerate}

\item Free Choices: This premise asserts that the choice made in
each region  as to which experiment will be performed in that
region can be treated as  a localized free variable.

\item No Backward in Time Influence: This premise asserts that
experimental  outcomes that have already occurred in an earlier
region (in frame LF) can  be considered to be fixed and settled
independently of which experiment  will be chosen and performed
later in a region spacelike separated from  the first.

\item Validity of Predictions of quantum theory (QT): Certain
predictions of quantum theory  in a Hardy-type experiment are valid.

\end{enumerate}

Then this theory or model violates the following Locality
Condition:  The free choice made in one region as to which
measurement will be performed  there has, within the theory, no
influence in a second region that is  spacelike separated from the
first. 

\section{Proof of the Theorem}

The theorem refers to the following Hardy-type\cite{5} experimental
set-up.  There are two experimental spacetime regions $L$ and $R$,
which are  spacelike separated, with $L$ lying earlier than $R$ in 
$LF$.  The experimenter in $L$ freely chooses either $L1$ or $L2$,
and an outcome, either $+$ or $-$, then appears in region $L$. Then
the  experimenter in region $R$ freely chooses either $R1$ or $R2$,
and one  or the other of the two alternative possible outcomes, $+$
or $-$,  then appears in $R$.
 
The detectors are assumed to be 100\% efficient, so that for
whichever measurement is chosen in $L$, one of the two alternative
possible outcomes of that measurement, either $+$ or $-$, will
appear in
$L$, with each of  these possibilities occurring about half the
time, and for whichever  measurement is then chosen in $R$, some
outcome of that measurement,  either $+$ or $-$,  will appear in
$R$.

For each of the two choices $L1$ or $L2$ available to the
experimenter  in $L$, and for each of the two alternative possible
outcomes $+$ or $-$  of that experiment, quantum theory makes
predictions for {\it both} of the  two alternative choices $R1$ and
$R2$ available to the experimenter in $R$.

In the statements that follow, the symbol $L1$ will be an
abbreviation of  the statement ``Experiment $L1$ is performed in
$L$.'' The symbols $L2$, $R1$, and $R2$ will have analogous
meanings.

The symbol $L1+$ will stand for the assertion, ``Experiment $L1$ is
performed  in $L$ and outcome $+$ of that experiment appears in
$L$.'' The symbols 
$L1-$, $L2+$, $L2-$, $R1+$, $R1-$, $R2+$, and $R2-$ have analogous
meanings. Using these abbreviations the first two pertinent predictions 
of QT for this Hardy setup are that:\cite{foot}

\smallskip \noindent Under the condition that $L2$ is performed in
$L$,
\begin{equation}
\mbox{If $R2+$, then $L2+$}
\end{equation}
and
\begin{equation}
\mbox{If ($L2+$ and $R1$) then $R1-$}.
\end{equation}
  
If, in accordance with our assumption, the choice made in $R$ does
not affect the outcome that has already occurred in $L$, then these
two  conditions entail:

\medskip \noindent {\bf Property 1}. Quantum theory predicts that if
an experiment  of the Hardy-type is performed then, $L2$ implies
$SR$, where,
\begin{eqnarray}
\mbox{$SR$ = If $R2$ is performed and gives outcome $+$, then if,
instead,} \nonumber \\
\mbox{$R1$ had been performed the outcome would have been $-$.}
\nonumber
\end{eqnarray}
\noindent {\bf Proof of Property 1}.
The concept ``instead'' is given an unambiguous meaning by the combination
of the premises of ``free choice,'' and ``no backward in time
influence;'' the choice between $R1$ and $R2$ is to be treated,
within the theory, as a free variable, and switching between $R1$
and $R2$ is required to leave any outcome in the earlier region $L$
undisturbed. But then statements (1) and (2) can be joined in tandem
to give the result $SR$.

The second two pertinent predictions of QT for this Hardy setup
are
\begin{eqnarray}
\mbox{Under the condition that $L1$ is performed in
$L$,} \nonumber \\
\mbox{If ($L1-$ and $R2$) then $R2+$}
\end{eqnarray}
and
\begin{equation}
\mbox{If ($L1-$ and $R1$) then sometimes $R1+$ .}
\end{equation}
If our premises are valid, then these two predictions entail:

\medskip \noindent {\bf Property 2}. Quantum theory predicts that if
an experiment  of the Hardy-type is performed then:
 ``$L1$ implies $SR$'' is false.

\medskip \noindent {\bf Proof of Property 2}:

Quantum theory predicts that if $L1$ is performed, then outcome $-$
appears about half the time. Thus if $L1$ is chosen, then there are 
cases where $L1-$ is true. But in a case where $L1-$ is true, the
prediction (3) asserts that the premise of $SR$ is true. But
statement (4), in conjunction with our two premises that give
meaning to ``instead,'' implies that the conclusion of $SR$ is not
true: if
$R1$ is performed instead of
$R2$, the outcome is not necessarily $R1-$, as it was in case $L2$.
So there  are cases where $L1$ is true but $SR$ is false. 

The conclusion is that in any theory or model in which the three
assumptions  (free choices, NBITI, certain predictions of quantum
theory) are valid, the  statement $SR$ must be always true if the
free choice in region $L$ is $L2$,  but must sometimes be false if
that free choice in $L$ is $L1$. But the truth  or falsity of $SR$
is defined by conditions on the truth or falsity of  statements
describing possible events located in region $R$. The fact that 
the truth of
$SR$ depends in this way on a free choice made  in region $L$,
which is space-like-separated from region $R$, can reasonably be
said to represent the existence within that theory or model of 
{\it some sort of\/} faster-than-light influence. This conclusion
is   discussed in the following sections.

\section{Hidden Variables and Counterfactual Arguments}

The previous argument  rests heavily on the use of
counterfactuals: the key statement $SR$ involves, in a situation in
which $R2$ is performed and gives outcome ``+,'' the idea ``if,
instead, $R1$ had been performed \ldots'' The Bell-type
hidden-variable assumptions also are  essentially counterfactual  in
nature. That is because the hidden variables identify instances in
which  the outcomes of all four of the alternative possible
experiments can be  considered simultaneously. But this similarity
does not mean that the three  assumtions used here are logically
equivalent to the Bell-type hidden-variable  assumptions. The
Bell-type assumptions, in their original deterministic
hidden-variable form, clearly entail the three assumptions used
here. But these three assumptions do not entail the Bell-type
hidden-variable assumptions. Indeed, our assumptions are, as
repeatedly emphasized, compatible with  orthodox Tomonaga-Schwinger
relativistic quantum field theory, whereas the  assumptions of
Bell, either in its original form, or in terms of the 
factorization property assumed in the later probabilistic form of
Bell's  assumption, \cite{2} are in direct logical conflict with
the principles of  relativistic quantum field theory. Thus the
assumptions made here are   logically weaker than those of the
traditional Bell's theorems.    

\section{Connection to Earlier Work}

The aim of this work is similiar to that of an earlier work of this
author.\cite{8}  That work was criticized by Unruh,\cite{9} 
Mermin,\cite{10} and  Shimony and  Stein\cite{11} on various
grounds. I have answered these objections.\cite{12, 13, 14}
However, the very existence of those challenges shows that the
approach used  in Ref.~\onlinecite{8} has serious problems, which
originate in the fact that it is based on  classical modal logic.

Classical modal logic provides the possibility of a concise
logically rigorous proof based on an established logic. However,
that virtue is overshadowed  by the following drawbacks:

\begin{enumerate}

\item Although the symbolic proof is concise and austere, that
brevity is based on a background that most physicists lack, which
means that most physicists  cannot fully understand it without a
significant investment of time.

\item The question arises as to whether the use of classical modal
logic begs the question by perhaps being based in implicit ways on
the deterministic notions of classical physics.

\item Classical modal logic itself is somewhat of an open question,
and it is not immediately clear to what extent these issues
undermine the proof.
\end{enumerate}

For these reasons I have in the present formulation
relied only on quantum thinking and language throughout: there is
no appeal to concepts unfamiliar to physicists.

But the present proof differs from the 1997 version by more than
just the use  of the language of physicists. The earlier proof
introduced the assumption  LOC2. That assumption was introduce in
order to set up a {\it reducio ad  absurdum} argument: it was meant
to be proved false. But a lot of the  discussion in my earlier
augument was whether I had adequately justified  this
``assumption,'' which, however, I was trying to prove false.  The
present version gives a straightforward proof of the key 
properties 1 and 2, without introducing the false assumption LOC2.

A subtle objection to the earlier argument was made by Shimony and
Stein\cite{11}:  Although the {\it explicit} condition for the truth
of
$SR$ is specified  entirely by the truth or falsity of statements
about possible events localized  in region $R$ --- and hence the
proven dependence of the truth of $SR$ upon  which experiment is
freely chosen in $L$ seems {\it explictly} to require an  influence
in $R$ of that choice made in  $L$ --- the word ``instead'' that 
occurs in
$SR$ harbors, they say, an {\it implicit} dependence of $SR$ on 
the choice made in $L$, and that the implicit dependence upon $L$
upsets  the conclusion. 

I believe this objection to be invalid. The word ``instead'' gets
its  meaning from the explicitly stated assumption that the choices
made by  the experimenters can be treated as free variables,
coupled with the explicitly stated assumption that, under any
condition chosen by the  experimenter in the earlier region $L$,
whatever he/she observes in that region can be considered to be
independent of the choice freely made in the later  region $R$ by
the other observer. These explicitly stated conditions are what 
enter into the logical structure of the proof, which then entails
that the  truth value of at least one statement whose truth or
falsity is specified  in terms of the outcome of a possible
experiment in region $R$ must depend  non-trivially on the choice
made by the experimenter in the spacelike  separated region $L$.
This conclusion represents {\it some sort of failure}  of the
notion that no influence {\it of any kind} can act over a
spacelike  interval.
 
\section{Connection to the Bohr-Einstein Debate}

The conclusion obtained here about faster-than-light influences
parallels and complements Bohr's reply to the paper of Einstein,
Podolsky, and Rosen.\cite{15} That discussion was not directly
about ``faster-than-light'' influences:  it was explicitly about
whether actions performed on one system can influence   ``real''
properties of another system. In the gedanken experiments 
described by EPR, the action of one experimenter could be made in
one place,  whereas the property in question could be measured far
away at the same time.  So the issue was {\it implicitly} about
the possibility of faster-than-light  influences of nearby actions
on faraway properties.

The EPR assumption was thus, essentially, that there was no
faster-than-light  influence of any kind. Bohr's response\cite{16}
was subtle. Rosenfeld's commentary\cite{17} on Bohr's response
makes it clear that Bohr's effort to compose a  response was not an
easy endeavor. 

The EPR argument rests strongly on counterfactual ideas, and it
stresses at  the end that ``Indeed, one would not arrive at our
conclusion if one  insisted that two or more physical quantities
can be regarded as simultaneous elements of reality {\it only when
they can be simultaneously measured or  predicted.}''\cite{15} 
(Italics are in the original.) But in spite of
this crucial use of counterfactual ideas by  EPR, Bohr did not
attack their argument on those grounds. I think it likely that Bohr
did not want to take the difficult road of trying to ban  all use
of counterfactual concepts in physics. Indeed, I believe he
recognized that counterfactual concepts do play an important role
in the pragmatic  approach to physics that he was pursuing. The
core of Bohr's response was  to insist that there was no
faster-than-light ``mechanical disturbance,'' but  to note that
``there is an influence on the very conditions that define  the
possible types of predictions regarding the future behavior of the 
system.'' 

A ``mechanical disturbance'' would be one capable of transmitting a
signal. The EPR argument does not prove, or attempt to prove, the
existence of any  nonlocal mechanical influence of this kind. Their
argument pertains to  more subtle kinds of considerations that
involve contemplating relationships  between experiments that
cannot be performed simultaneously, but that have  outcomes that
can be predicted with certainty (that is, with probability unity), 
and with what ensues from trying to impose upon these considerations
also  the idea that any predicted property that can be considered
``real'' in one  region cannot depend upon what a faraway
experimenter freely chooses to do.  The argument of EPR is
therefore of the same general kind as the one given  here. On the
other hand, the EPR argument is very different from arguments  of
the Bell type, because the existence of hidden-variables was a
putative  {\it consequence} of the EPR argument, whereas it is an
{\it assumption}  of the Bell-type local hidden-variable (or local
realism) theorems. 

Bohr challenged not the counterfactual aspect of the EPR argument, 
but rather the strong form of the locality assumption assumed by
EPR:  he claimed that ``there is an influence (of the nearby
choice) upon the  very conditions that define the possible types of
predictions regarding  the future behavior of the (faraway)
system.'' 

That statement by Bohr, the core of his reply to EPR, has puzzled many
philosophers and physicists, and Bohr himself\cite{18} recognized
its difficulty.  The present argument can be considered to be an
effort to further explicate  the nature of the subtle quantum
non-locality that Bohr apparently recognized,  and whose existence
falsified, in his opinion, the EPR argument for the  incompleteness
of quantum theory, that is, falsified, {\it within the conceptual
framework of Copenhagen quantum theory}, the EPR argument for  the
existence of local hidden variables of the kind postulated in the
Bell  local hidden-variable (or local-realism) theorems. The
present theorem separates the {\it locality} aspect of the Bell
local-realism postulate from the {\it reality} aspect by imposing a
{\it valid} (according to  Copenhagen thinking) (effectively
one-way) locality condition that is  expressed exclusively in terms
of observables, and that is true in  relativistic quantum field
theory, and then deducing the failure of a 
no-faster-than-light-influence condition in the other directions,
rather  than by proving the failure of a strong, and very alien to
quantum thinking,  property that intricately combines the locality
condition with a ``realism'' assumption that conflicts with quantum
philosophy.

This argument has been carried out in a purely theoretical plane.
It rests on a highly idealized theoretical conception of the
experimental conditions and on purely theoretical assumptions. No
attempt is made here to derive  conclusions directly from empirical
data, as certain approaches to Bell's  theorem seek to do. A recent
discussion of some of the difficulties that ensue  if one tries to
argue directly from experimental data is given in
Ref.~\onlinecite{19}.

\appendix

\section{Appendix: Connection to Hardy's notation} 
To obtain these four predictions from Hardy's paper,\cite{5} one
transforms my notation into Hardy's using 
$$
L1(+,-)\rightarrow (c_1,d_1)
$$
$$
L2(+,-)\rightarrow (v_1,u_1)
$$
$$
R1(+,-)\rightarrow (d_2,c_2)
$$
$$
R2(+,-)\rightarrow (u_2,v_2)
$$
and uses the three zero's connecting my pairs of states $(R2+,L2-)$,
$(L2+,R1+)$, and $(L1-,R2-)$ that arise from his Eq.~(13), 
respectively to obtain my statements (1), (2), and  (3),
respectively,  and uses his Eq.~(13.d), which says that my matrix
element $(L1-,R1+)$  is positive, to obtain my statement (2.4).

\begin{acknowledgements}
Extensive discussions with Asher Peres, Philippe Eberhard, Jerry
Finkelstein, and particularly Abner Shimony, have contributed
importantly to the  development of this paper. This work was
supported in part by the Director, Office of Science,  Office of
High Energy and Nuclear Physics, of the U.S.\ Department of Energy 
under Contract DE-AC03-76SF00098.
\end{acknowledgements}

[xx need titles of articles! xx]

\end{document}